# Tunable Doping in Hydrogenated Single Layered Molybdenum Disulfide

Debora Pierucci[1], Hugo Henck[1], Zeineb Ben Aziza[1], Carl H. Naylor[2], A. Balan[2], Julien E. Rault[3], M. G. Silly[3], Yannick J. Dappe[4], François Bertran[3], Patrick Le Fevre[3], F. Sirotti[3], A.T Charlie Johnson[2] and Abdelkarim Ouerghi[1*]

[1]Centre de Nanosciences et de Nanotechnologies, CNRS, Univ. Paris-Sud, Université Paris-Saclay, C2N – Marcoussis, 91460 Marcoussis, France
[2]Department of Physics and Astronomy, University of Pennsylvania, 209S 33rd Street, Philadelphia, Pennsylvania 19104, USA
[3] Synchrotron-SOLEIL, Saint-Aubin, BP48, F91192 Gif sur Yvette Cedex, France
[4] SPEC, CEA, CNRS, Université Paris-Saclay, CEA Saclay, 91191 Gif-sur-Yvette Cedex, France

**Abstract:** Structural defects in the molybdenum disulfide ($MoS_2$) monolayer are widely known for strongly altering its properties. Therefore, a deep understanding of these structural defects and how they affect $MoS_2$ electronic properties is of fundamental importance. Here, we report on the incorporation of atomic hydrogen in mono-layered $MoS_2$ to tune its structural defects. We demonstrate that the electronic properties of single layer $MoS_2$ can be tuned from the intrinsic electron (n) to hole (p) doping *via* controlled exposure to atomic hydrogen at room temperature. Moreover, this hydrogenation process represents a viable technique to completely saturate the sulfur vacancies present in the $MoS_2$ flakes. The successful incorporation of hydrogen in $MoS_2$ leads to the modification of the electronic properties as evidenced by high resolution X-ray photoemission spectroscopy and density functional theory calculations. Micro-Raman spectroscopy and angle resolved photoemission spectroscopy measurements show the high quality of the hydrogenated $MoS_2$ confirming the efficiency of our hydrogenation process. These results demonstrate that the $MoS_2$ hydrogenation could be a significant and efficient way to achieve tunable doping of transition metal dichalcogenides (TMD) materials with non-TMD elements.

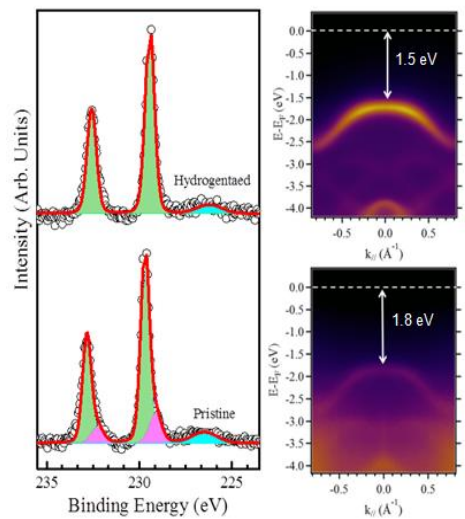

Following similar trend as graphene, the two-dimensional (2D) transition metal dichalcogenides ($MX_2$) have attracted much interest recently thanks to their versatile electronic and optical properties. However, unlike graphene that does not have a bandgap, they exhibit attractive properties such as indirect-to-direct bandgap crossover with decreasing number of atomic layers and strong photoresponses.[1–3] Moreover, 2D $MX_2$ materials have interesting electronic structure *i.e.* their tunable bandgap by varying the layer thickness and strain,[4,5] their edge-dependent semiconducting-to-metallic transitions,[6] in addition to the transformations between different S-Mo-S atoms stacking geometries (for instance, 2H to 1T phase of monolayer $MoS_2$[7]). In contrast to graphene where only small bandgap (few hundred meV) can be opened by strain and other methods,[8] large bandgap tunability can be obtained with 2D $MX_2$ where we can switch from a semiconducting (with a few eV bandgap) to a metallic form.[9] This large tunability allows broadening the applications of $MX_2$ in nanoelectronic devices.

Furthermore, the presence of defects can induce deep gap states responsible for the intrinsic doping of these materials.[10,11] For instance, in the case of $MoS_2$, sulfur mono-vacancies represent the defects with the lowest formation energy[12–14] and the most common generally present in $MoS_2$ flakes.[12] In particular, these sulfur vacancies (Sv) cause the presence of unsaturated electrons in the surrounding Mo atoms and act as electrons donors[15,16] responsible for the n-type doping of $MoS_2$.[17] It has already been demonstrated that the intercalation of oxygen,[18] nitrogen,[19] or niobium[20] can reduce the intrinsic doping of $MoS_2$. Recent experiments using mainly Raman spectroscopy and high resolution X-ray photoemission spectroscopy (HR-XPS) demonstrated that the n doped $MoS_2$ can be tuned to p doped by $N_2$ plasma.[21] The introduction of dopants in $MoS_2$ lattice represents a potential route to achieve stable $MoS_2$ with different functionalities. Therefore, it is crucial to develop controllable techniques to make possible the tunability of $MoS_2$ properties without degrading the quality of $MoS_2$ flakes.

In this work we report an easy and effective chemical doping method by hydrogen atoms, already used as dopants for graphene[22,23] and we show that hydrogen represents an effective way to passivate sulfur vacancies. We study the electronic properties of the hydrogen-doped $MoS_2$, by exposing the sample to atomic hydrogen gas. HR-XPS studies reveal the n type doping of pristine single layer $MoS_2$ which is also confirmed by the measurement of the valence band maximum (VBM). We demonstrate that the atomic hydrogen doping can tune this intrinsic n doping until reaching p doped $MoS_2$. The analysis of the total density of states (LDOS) of the hydrogenated $MoS_2$ by first principle calculations using density functional theory (DFT) shows the presence of new gap states close to the VBM, confirming the p-type behavior of the hydrogenated $MoS_2$ on graphene. We also checked the quality of the hydrogenated-$MoS_2$ using angle resolved photoemission spectroscopy (ARPES) measurements, which show sharp band structures. These results confirm the validity of the doping process without a degradation of the $MoS_2$ flakes.

**Results and discussions:**

We used p-type graphene substrate obtained by hydrogenation of epitaxial graphene. This bilayer graphene/SiC(0001) presents several advantages. The graphene/SiC is conducting therefore particularly suited for XPS/ARPES measurements[24]. Also, the graphene presents a chemically inert surface due to the strong coupling of all its $p_z$ atomic orbitals which are stabilized in a giant delocalized π bonding system.[25,26] This configuration inhibits the possibility of covalent addition and then the interaction with atomic hydrogen. Moreover, the pre-hydrogenation treatment of the graphene/SiC substrate,[27] ensures the saturation of the dangling bonds present at the SiC/buffer layer interface,[22] preventing the hydrogen intercalation in graphene during the MoS$_2$ treatment. Large (20-100 μm) MoS$_2$ flakes are grown by chemical vapor deposition (CVD) and then further deposited on the p doped bilayer graphene substrate using wet transfer process.[28] To further clean the surface and interface of the MoS$_2$/graphene heterostructure, we annealed the samples at T = 300°C for 30 min in ultra-high vacuum (UHV) (P ≈ 10$^{-10}$ mbar).[29] More details about the growth and sample preparation are given in the method section.

After the transfer, the quality of the MoS$_2$ flakes was verified by Raman spectroscopy. The two Raman maps in Figure 1(a) and (b) show the intensity of the E$^1_{2g}$ and A$_{1g}$ Raman modes, respectively, of the MoS$_2$ on graphene layer. The two Raman peaks correspond to the in-plane vibrations (E$^1_{2g}$) and out of plane vibrations (A$_{1g}$) of Mo and S atoms in the MoS$_2$. These maps reveal a large and uniform MoS$_2$ flake with only some variations of contrast in the whole flake. This uniformity of the peaks intensities indicates the good quality of our MoS$_2$ flakes. In order to investigate the effects of atomic hydrogen exposures to single layer MoS$_2$ on p doped bilayer graphene/SiC, the sample was exposed to several atomic hydrogen doses (~ 5 × 10$^2$ L, 5 × 10$^3$ L and 8 × 10$^4$ L, where 1 Langmuir (L) = 10$^{-6}$ Torr × s). The molecular di-hydrogen was cracked by a hot tungsten filament (1400°C, approximately 5 cm from the sample), as illustrated in the schematic of Figure 1(c). Figure 1(d) shows an optical micrograph of a representative single layer flake on graphene/SiC after the hydrogenation process.

To examine the atomic composition as well as the chemical bonding environment of our samples, HR-XPS measurements at room temperature were carried out for pristine and hydrogen doped MoS$_2$. The spectra were collected in a surface sensitive condition at a photon energy at hν = 340 eV (the photoelectrons were detected at 0° from the sample surface normal $\vec{n}$ and at 46° from the polarization vector $\vec{E}$ ). The C-1s spectrum corresponding to the underlayer p doped graphene is shown in Figure S1. The experimental data points are displayed as dots. The solid line is the envelope of fitted components. The pristine underlayer C1s spectra consist of two components located at the binding energies of 282.5 and 284.3 eV. These components correspond to bulk SiC and bilayer graphene, respectively.[27] The shape and the position of the C 1s peak, arising from the graphene and the SiC underlayer, have not changed after the exposition to three hydrogen doses, with respect to the pristine one. This confirms that the hydrogen did not interact with the graphene/SiC underlayer. The evolution of the Mo 3d and S 2p peaks from the pristine MoS$_2$ through the three different hydrogen doses are shown in Figure 2(a) and (b), respectively. For the pristine MoS$_2$, the Mo 3d spectrum contains one main doublet component at binding energy (BE) Mo 3d$_{5/2}$ = 229.7 eV (3d$_{5/2}$:3d$_{3/2}$ ratio of 0.66 and a spin-orbit splitting of 3.10 eV[30] ) related to a Mo$^{4+}$ in a sulfur environment[31] with a trigonal prismatic phase (1H-MoS$_2$). A smaller contribution visible at lower BE (~ -0.52 eV) with respect to this main doublet peaks is the signature of a defective/sub-stoichiometric MoS$_2$ with Sv.[12,31] The weight of this component (between 15-18% of the whole Mo 3d spectrum) is not representative of a single MoS$_2$ flake due to the large X-ray beam size (~ 100-150 μm diameter), but it gives information on the percentage of defective MoS$_2$ in the explored area. The additional peak at BE = 226.5 eV is due to the sulfur 2s peak. The S 2p spectrum of the pristine MoS$_2$ present only one main doublet at BE S 2p$_{3/2}$ = 162. 5 eV (2p$_{1/2}$:2p$_{3/2}$ ratio of 0.5 and a spin-orbit splitting of 1.19 eV[30]) as expected for divalent sulfide ions (S$^{2-}$) in 1H-MoS$_2$.[31,32] These BE values for the Mo 3d and S 2p indicate an intrinsic n-type doping of the MoS$_2$ flakes,[15] mostly induced by the Sv which act as an electron donating defects. No other components are present on the Mo 3d and S 2p spectra related to oxygen or carbon bonds[33–35] indicating that no contaminations are present on the sample and confirming the high quality of this heterostructure.

After the first exposure to atomic hydrogen, the core level peaks of Mo 3d and S 2p show a rigid shift of about 0.1 eV toward lower binding energies. The peaks related to Sv (highlighted in pink color in Figure 2(a)) have decreased compared to those of the pristine $MoS_2$ (18-15% to 13-10% respectively). This can be explained by the partial saturation of these vacancies with hydrogen atoms forming Mo-H bonds. These Sv peaks completely disappear after the second dose and all core levels BE are shifted by -0.3 eV, indicating a complete saturation of the Sv. In this case both, the Mo 3d and S 2p peaks became sharper (FWHM = 0.6 to 0.5 eV). The saturation of these vacancies gradually reduced the n-type doping of the $MoS_2$ decreasing the distance of the VBM to the Fermi level (FL) as shown in Figure 3(a) and (b), from 1.25 eV for the pristine $MoS_2$ to 1.05 eV (almost mid-gap, considering a quasi-particle bandgap of 2 eV[36]) for the second hydrogen dose. We notice that the spectra in Figure 3(a) show similar characteristic peaks, before and after hydrogenation (for different doses), that are related to $MoS_2$.[37] The shifts of hydrogenated $MoS_2$ shown in Figure 3(b) are caused by the rigid energy shift of all core-levels. When the hydrogen dose is further increased the core level peaks still shift toward lower binding energy (-0.5 eV), which correspond to a distance of the VBM to the FL of about 0.75 eV. This means that a p-type doping is now induced in the $MoS_2$ flakes. This huge hydrogen dose ($\sim 8 \times 10^4$ L) probably starts to induce new Sv, directly replaced by hydrogen forming Mo-H bonds.[38–40] The formations of Mo-H bonds and the saturation of the Sv replace the donor gap state present in the pristine $MoS_2$ to new acceptor gap state related to the Mo-H bond formation as shown below by DFT calculations. Moreover, the Mo 3d and S 2p spectra show also the presence of a new component at lower binding energy with respect to the main doublet peaks ($\sim 0.8$ eV). These components are probably related to the formation of the octahedral 1T phase of $MoS_2$ induced by the high density of hydrogen atoms present in the sample.

The electronic structure and the stability of the hydrogenated $MoS_2$, kept one month under ambient conditions after the hydrogenation (a dose of $5 \times 10^3$ L), were studied using ARPES (Figure 4). Figures 4(a) and (b) show the ARPES momentum energy image, before and after hydrogenation respectively, around the Γ point of the Brillouin zone (BZ). The expected signal of the monolayer $MoS_2$ valence band is visible.[41,42] The shape of the upper band of the valence bands is not modified but only shifted toward the FL because of the exposure to the hydrogen atoms. Moreover, we notice the same shift for the other bands around -3 eV and -4 eV (highlighted by the white circles). Hence, we can claim that the hydrogenation induces a uniform shift of all the bands forming the valence bandstructure of $MoS_2$. The sharp and intense experimental band structure of Figure 4(b) confirms the hydrogenation process preserves the high structural quality of the $MoS_2$. The ARPES quality is even better after hydrogenation suggesting that the hydrogen treatment clean the surface of the $MoS_2$.

ARPES measurement, perpendicular to the Γ–K direction of the p-doped graphene underlayer BZ before and after the hydrogenation of $MoS_2$/p-doped graphene underlayer, are shown in the supplementary information in Figure S2. In this intensity map, the graphene underlayer presents two π bands, which is the clear signature of the interlayer decoupling induced by the hydrogenation of the epitaxial graphene giving rise to a quasi-free standing bilayer graphene and a completely saturated SiC surface. Moreover, in this sample orientation, the two branches of the graphene π bands can be observed[43] allowing the determination of the Dirac point position with respect to the FL. Close inspection of the dispersion relation around K reveals that the FL is located at 0.25 eV below the Dirac points for the pristine MoS$_2$/graphene heterostructure and is not modified after the hydrogen dose. This value corresponds to a hole doping level of the graphene underlayer of about $3.8 \times 10^{12}$ cm$^{-2}$.

In order to probe the effects of atomic hydrogen on the electronic properties of hydrogen doped MoS$_2$ ($8 \times 10^4$ L), we performed micro-Raman spectroscopy on two flakes with the same orientation angle with respect to the graphene underlying substrate. This choice is made so that we can assume that the flakes are subjected to the same orientation-induced strain.[36,44,45] In Figure 4(c), the Raman spectra of MoS$_2$ before and after the hydrogenation are shown. The two vibrational modes are separated by about 19 cm$^{-1}$ which corresponds to a single layered MoS$_2$ as was reported for MoS$_2$ monolayer.[46,47] After the treatment, we clearly observe that a downshift of the E$^1_{2g}$ peak has occurred together with a splitting of this peak. This is likely due to a strain induced by the hydrogen intercalation (which is probably responsible for the appearance of the 1T-phase);[48] this strain is tensile as indicated by the direction of the E$^1_{2g}$ shift.[45,49–51] The peak splitting gave rise to two distinctive E$^1_{2g}$ modes, similarly to what was observed by A. Azcatl *et al.* who doped the MoS$_2$ using nitrogen plasma.[21] Furthermore, the A$_{1g}$ peak is slightly upshifted which can be explained by a change of the doping level in MoS$_2$[36,52] *i.e.* the saturation of the Sv, the hydrogen atoms introduction, and eventually the removal of S atoms.[14,53] On the other hand, the increase of the A$_{1g}$ peak intensity or in other words the increase of the ratio of intensities A$_{1g}$/E$^1_{2g}$ is consistent with what was reported on the p doping of MoS$_2$.[20] According to these Raman results, we show that the hydrogenation of MoS$_2$ was successfully done leading to a non-defective MoS$_2$. Photoluminescence (PL) spectra of pristine and hydrogenated MoS$_2$ are provided in Figure S3. We notice that the PL intensity decreases after hydrogenation; this PL quenching can be explained by the suppression of defects from MoS$_2$ at room temperature as was observed for pristine and irradiated MoS$_2$.[14] This confirms again that the used process is noninvasive and that it is possible to hydrogenate MoS$_2$ even with better quality.

In order to gain insights into this doping change from n to p of the MoS$_2$ flakes we performed first principle calculations based on DFT. Figure 5 shows the total density of state (DOS) of a perfect MoS$_2$ monolayer (black curve), MoS$_2$ with Sv (red curve) and Sv saturated by hydrogen atoms (blue curve). As expected, the presence of Sv induces defect state in the gap, located at about 0.75 eV below the conduction band minimum (CBM). This donor state is in fact responsible for the intrinsic n-type doping of the pristine MoS$_2$ flake as measured by XPS. Our calculation shows that Mo-H bonds induce new gap state at about 0.16 eV above the valence band maximum (VBM). The presence of this acceptor like state confirms the p-type behavior observed experimentally after the MoS$_2$ hydrogenation. The combination of XPS experiments and DFT gives a global picture of the structural and electronic MoS$_2$ properties after the hydrogenation process. A single layer MoS$_2$ includes well known Sv as displayed in Figure 6(a). After the first hydrogenation (Figure 6(b)), the hydrogen atoms react with the unbound Mo atoms. This model explains the elimination of defect peaks in Mo 3d. At high atomic hydrogen dose, if new Sv vacancies are created by high hydrogen dose, they are readily replaced by the hydrogen atoms increasing the p-type doping of the MoS$_2$ as indicated by the energy shift of the Mo-3d and S-2p core level spectra toward lower binding energy and the reduction of the energy distance of the VBM with respect to the FL upon hydrogenation (Figure 6(c)).

**Conclusions:**

In summary, we conducted a comprehensive study of the interaction of atomic hydrogen with monolayer MoS$_2$. The pristine MoS$_2$ has sulfur vacancies, which are responsible for its intrinsic n-type doping. During hydrogen dosing, the hydrogen atoms react with the under-coordinated Mo atoms filling the Sv. When all the vacancies are filled, the MoS$_2$ becomes almost intrinsic (VBM = 1.05 eV). At higher atomic hydrogenation, the hydrogen breaks the Mo-S bonds and passivates the new created vacancies inducing a p-type doping in the MoS$_2$ sample. This model suggested by the HR-XPS results is corroborated by DFT calculations predicting the presence of an acceptor gap state upon the formation of the Mo-H bonds. As Raman and ARPES measurements show that the hydrogen dosing process does not alter the quality of the MoS$_2$ flakes since it preserves a well-defined electronic structure. Therefore, we see this hydrogenation process as an efficient method to reduce the defects in MoS$_2$ flakes, control, and switch from n to p the intrinsic as-grown doping. Such control of doping in 2D TMDs is of great importance for potential nanoscale, flexible devices applications based on these materials.

**Methods:**

**Bilayer graphene on SiC(0001):** Bilayer graphene was produced by thermal heating the SiC(0001) substrate. Before the graphitization, the substrate was etched with hydrogen (100% H$_2$) at 1550 °C to produce well-ordered atomic terraces of SiC. The substrate was heated to 1000 °C and then further heated at 1550 °C under argon (800 mbar).

**Growth and transfer of MoS$_2$:** Large scale MoS$_2$ monolayer flakes (≈20 to ≈100 μm) have been grown by Chemical Vapour Deposition (CVD) on oxidized silicon substrate (see methods and ref [54]). The MoS$_2$ flakes transferred onto the graphene retain their triangular shapes with unchanged lateral sizes. Before any measurement, the MoS$_2$ sample was annealed at 300 °C for 30 min in ultra-high vacuum (P ≈ 10$^{-10}$ mbar), in order to remove the residual surface contaminations induced by the wet transfer.

**Atomic hydrogen doping:** The hydrogenation process has been performed at room temperature under a pressure of 2 × 10$^{-5}$ mbar of H$_2$ in order to avoid hydrogen etching of the MoS$_2$ layer. The hydrogenation process lasted 3 hours, under a lower pressure of H$_2$, in order to control the phenomena that may occur on the MoS$_2$ properties. This procedure has been repeated twice, after each the sample was fully characterized by means of XPS measurements.

**HR-XPS:** HR-XPS experiments were carried out on the TEMPO beamline[55] (SOLEIL French synchrotron facility) at room temperature. The photon source was a HU80 Apple II undulator set to deliver horizontally linearly polarized light. The photon energy was selected using a high-resolution plane grating monochromator, with a resolving power E/ΔE that can reach 15,000 on the whole energy range (45 - 1500 eV). During the XPS measurements, the photoelectrons were detected at 0° from the sample surface normal $\vec{n}$ and at 46° from the polarization vector $\vec{E}$. The spot size was 80 × 40 (H×V) μm$^2$.

A Shirley background was subtracted in all core level spectra. The C 1s spectrum was fitted by a sum of a Gaussian function convoluted with a Doniach-Sunjic lineshape. An asymmetry factor α was used, where α = 0.1 eV (peak G) and α = 0 eV (SiC). The Mo 3d and S 2p spectra were fitted by sums of Voigt curves, *i.e*, the convolution of a Gaussian (of full-width at half-maximum GW) by a Lorentzian (of full-width at half-maximum LW). The LW was fixed at 90 meV for Mo 3d and S 2p.[30]

**Angle-resolved Photoemission spectroscopy:** The ARPES measurements were conducted at the CASSIOPEE beamline of Synchrotron SOLEIL (Saint-Aubin, France). We used horizontally linearly polarized photons of 50 eV and a hemispherical electron analyzer with vertical slits to allow band mapping. The total angle and energy resolutions were 0.25° and 25 meV. The mean diameter of the incident photon beam was smaller than 50 $\mu$m. All ARPES experiments were done at room temperature.

**DFT calculations:** First-principles calculations have been performed using a very efficient DFT localized orbital molecular dynamic technique (FIREBALL).[56,57] Basis sets of sp3d5 for S and Mo were used with cutoff radii (in atomic units) s = 3.9, p = 4.5, d = 5.0 (S) and s = 5.0, p = 4.5, d = 4.8 (Mo).[45] In this study we have considered a standard (5x5) unit cell of $MoS_2$ that has been optimized as a pristine, with an S vacancy and with an H-filled S vacancy, before calculating the corresponding DOS.

**Supporting Information:**

The Supporting Information is available free of charge on the ACS Publications website at DOI:

Supplementary figures: Figure S1: High-resolution C 1s spectrum (hv = 340 eV) of pristine graphene and graphene after three differ hydrogen doses ((~ 5 × 10$^2$ L, 5 × 10$^3$ L and 8 × 10$^4$ L). Figure S2: a) and b) ARPES at room temperature of $MoS_2$/graphene and hydrogenated $MoS_2$/graphene heterostructure, measured at hv = 100 eV and hv = 50 eV respectively, around the K-point and along the graphene ΓK direction. Figure S3: Photoluminescence spectra of pristine and hydrogenated $MoS_2$.

**Acknowledgements:** This work was supported by the ANR H2DH grants. C.H.N. and A.T.C.J. Acknowledge support from the National Science Foundation EFRI-2DARE program, grant number ENG- 1542879.

**Competing financial interests:** The authors declare no competing financial interests.

**Figure captions:**

**Figure 1:** a) and b) Micro-Raman maps of $E^1_{2g}$ and $A_{1g}$ peaks of the $MoS_2$ on epitaxial graphene; c) Schematic illustration of our current method for hydrogen-doped $MoS_2$; d) Typical optical image of a hydrogenated $MoS_2$ flake on epitaxial graphene/SiC.

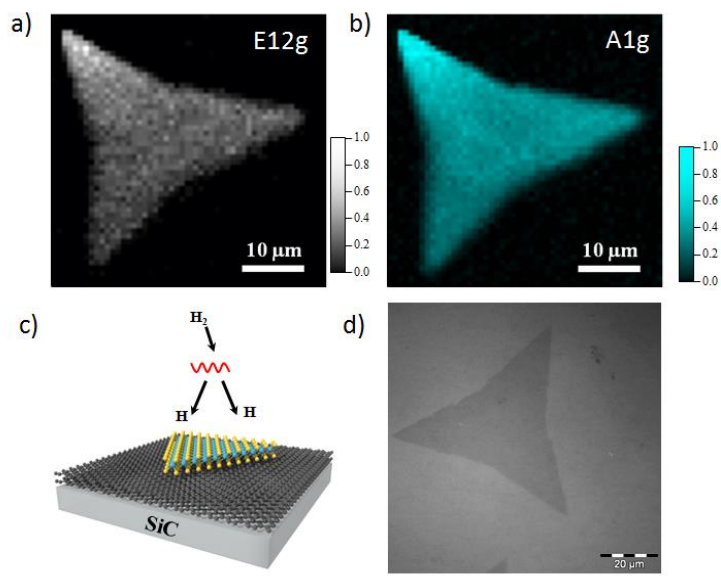

Figure 2: XPS measurement for MoS$_2$ at different doses of atomic hydrogen: a) Mo-3d core level, b) S-2p core level at hν = 340 eV. The experimental data points are displayed as dots. The solid line is the envelope of fitted components.

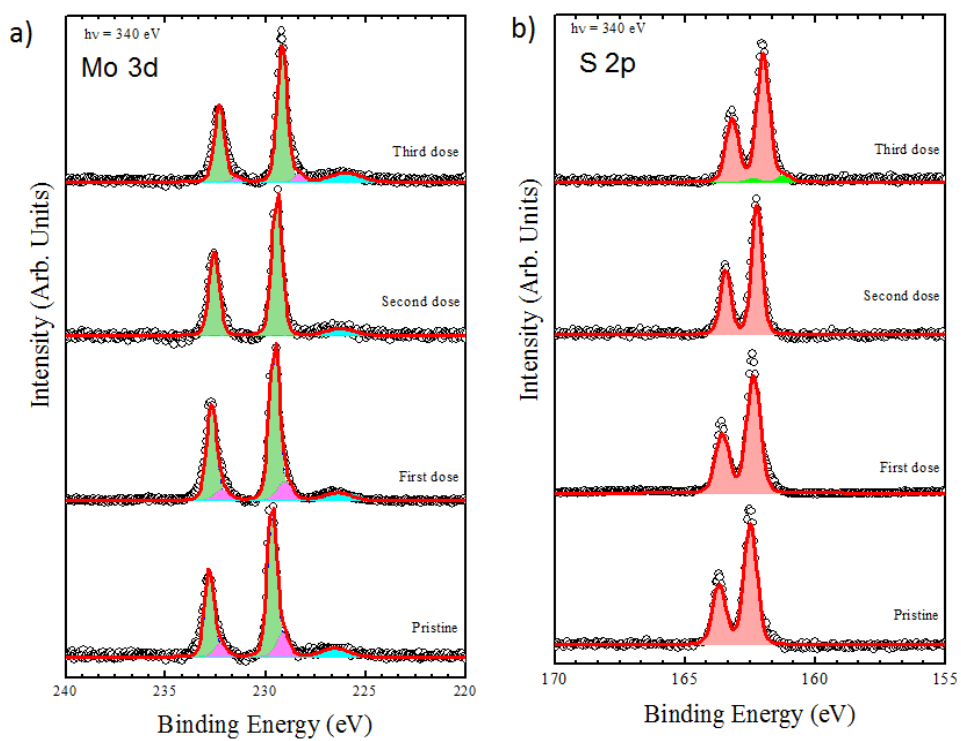

**Figure 3:** a) ARPES intensity integrated spectra as a function of the binding energy of pristine MoS$_2$ and hydrogenated MoS$_2$ (for three doses) and b) Zoom of a) . The experimental data points are displayed as dots.

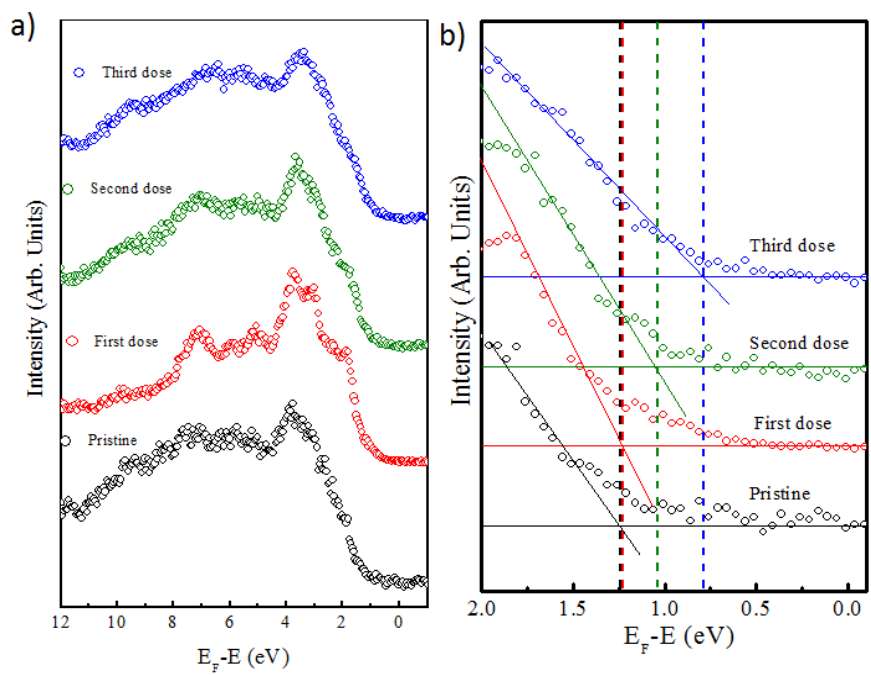

**Figure 4:** a) and b) ARPES measurements at room temperature of pristine MoS$_2$/graphene at hν = 50 eV and hydrogenated MoS$_2$/graphene heterostructure at hν = 100 eV, respectively along the graphene ΓK direction. The Fermi level position is located at the zero of the binding energy (marked as a white dashed line), the white circles highlight the shift in the position of the band located at around -4 eV; c) Comparison between the Raman spectra of MoS$_2$ before (in black line) and after (in blue line) hydrogenation.

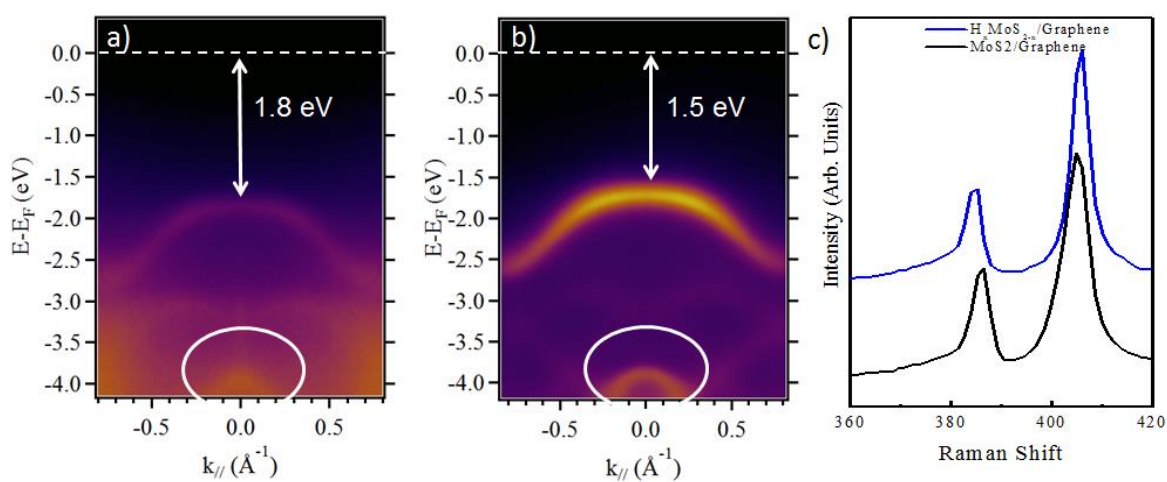

**Figure 5:** The total DOS of pristine MoS$_2$ (black line), MoS$_2$ with Sv (red line), and hydrogenated MoS$_2$ (blue line) using DFT calculations.

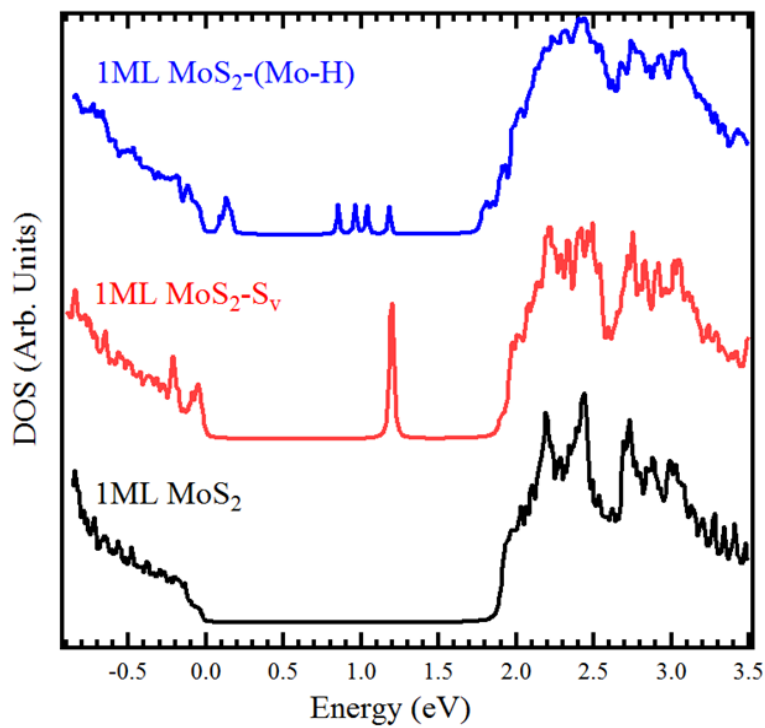

**Figure 6:** Band diagram showing the evolution of MoS$_2$ Fermi level: a) before hydrogenation, b) upon the first exposure to hydrogen, and c) after the complete hydrogenation.

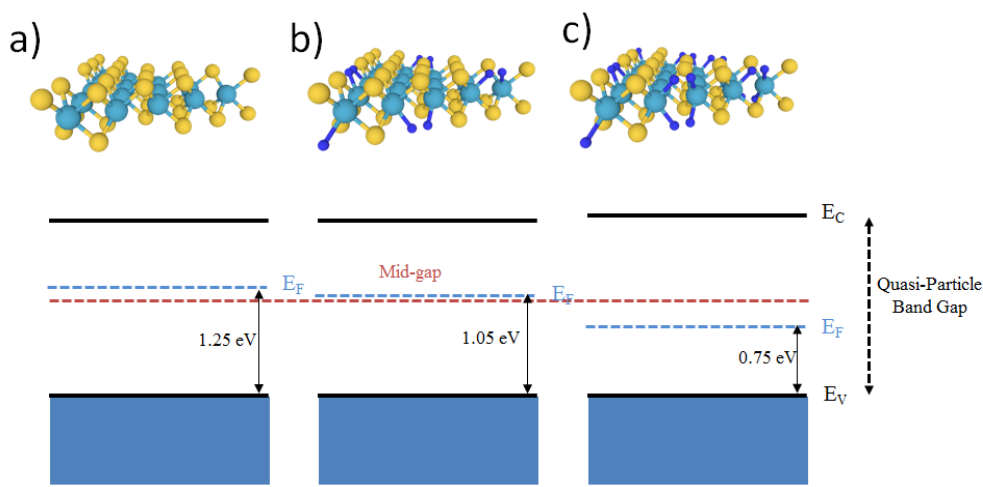

**References**

(1) Furchi, M. M.; Polyushkin, D. K.; Pospischil, A.; Mueller, T. Mechanisms of Photoconductivity in Atomically Thin MoS2. *Nano Lett.* **2014**, *14* (11), 6165–6170.

(2) Buscema, M.; Island, J. O.; Groenendijk, D. J.; Blanter, S. I.; Steele, G. a; van der Zant, H. S. J.; Castellanos-Gomez, A. Photocurrent Generation with Two-Dimensional van Der Waals Semiconductors. *Chem. Soc. Rev.* **2015**, *44* (11), 3691–3718.

(3)  Lopez-Sanchez, O.; Lembke, D.; Kayci, M.; Radenovic, A.; Kis, A. Ultrasensitive Photodetectors Based on Monolayer MoS2. *Nat. Nanotechnol.* **2013**, *8* (7), 497–501.

(4)  Splendiani, A.; Sun, L.; Zhang, Y.; Li, T.; Kim, J.; Chim, C. Y.; Galli, G.; Wang, F. Emerging Photoluminescence in Monolayer MoS2. *Nano Lett.* **2010**, *10* (4), 1271–1275.

(5)  Conley, H. J.; Wang, B.; Ziegler, J. I.; Haglund, R. F.; Pantelides, S. T.; Bolotin, K. I. Bandgap Engineering of Strained Monolayer and Bilayer MoS 2. *Nano Lett.* **2013**, *13* (8), 3626–3630.

(6)  Li, Y.; Tongay, S.; Yue, Q.; Kang, J.; Wu, J.; Li, J.; Li, Y.; Tongay, S.; Yue, Q.; Kang, J.; Wu, J.; Li, J. Metal to Semiconductor Transition in Metallic Transition Metal Dichalcogenides. *Appl. Phys. Lett.* **2013**, *114*, 174307.

(7)  Tang, Q.; Jiang, D. Stabilization and Band-Gap Tuning of the 1T-MoS 2 Monolayer by Covalent Functionalization. *Chem. Mater.* **2015**, *27*, 3743–3748.

(8)  Pierucci, D.; Sediri, H.; Hajlaoui, M.; Velez-fort, E.; Dappe, Y. J.; Silly, M. G.; Belkhou, R.; Shukla, A.; Sirotti, F.; Gogneau, N.; Ouerghi, A. Self-Organized Metal – Semiconductor Epitaxial Graphene Layer on off-Axis 4H-SiC ( 0001 ). *Nano Res.* **2015**, *8* (3), 1026–1037.

(9)  Li, Y.; Duerloo, K.-A. N.; Wauson, K.; Reed, E. J. Structural Semiconductor-to-Semimetal Phase Transition in Two-Dimensional Materials Induced by Electrostatic Gating. *Nat. Commun.* **2016**, *7*, 10671.

(10) Krivosheeva, A. V; Shaposhnikov, V. L.; Borisenko, V. E.; Lazzari, J. Theoretical Study of Defect Impact on Two-Dimensional MoS 2. *J. Semicond.* **2015**, *36* (12), 122002.

(11) Lin, Z.; Carvalho, B. R.; Kahn, E.; Lv, R.; Rao, R.; Terrones, H. Defect Engineering of Two-Dimensional Transition Metal Dichalcogenides. *2D Mater.* **2016**, *3*, 22002.

(12) Zhou, W.; Zou, X.; Najmaei, S.; Liu, Z.; Shi, Y.; Kong, J.; Lou, J. Intrinsic Structural Defects in Monolayer Molybdenum Disul Fi de. *Nano Lett.* **2013**, *13* (6), 2615–2622.

(13) Liu, D.; Guo, Y.; Fang, L.; Robertson, J.; Liu, D.; Guo, Y.; Fang, L.; Robertson, J. Sulfur Vacancies in Monolayer MoS2 and Its Electrical Contacts. *Appl. Phys. Lett.* **2013**, *103*, 183113.

(14) Parkin, W. M.; Balan, A.; Liang, L.; Das, P. M.; Lamparski, M.; Carl, H.; Rodríguez-manzo, J. A.; Johnson, A. T. C.; Meunier, V.; Drndic, M. Raman Shifts in Electron-Irradiated Monolayer MoS 2. *ACS Nano* **2016**, *10* (4), 4134–4142.

(15) Addou, R.; Mcdonnell, S.; Barrera, D.; Guo, Z.; Azcatl, A.; Wang, J.; Zhu, H.; Hinkle, C. L.; Quevedo-lopez, M.; Alshareef, H. N.; Colombo, L.; Hsu, J. W. P.; Wallace, R. M. Impurities and Electronic Property Variations of Natural MoS 2 Crystal Surfaces. *ACS Nano* **2015**, No. 9, 9124–9133.

(16) González, C.; Biel, B.; Dappe, Y. J. Theoretical Characterisation of Point Defects on a MoS2 Monolayer by Scanning Tunnelling Microscopy. *Nanotechnology* **2016**, *27*, 105702.

(17) Guo, Y.; Liu, D.; Robertson, J. Chalcogen Vacancies in Monolayer Transition Metal Dichalcogenides and Fermi Level Pinning at Contacts. *Appl. Phys. Lett.* **2015**, *106* (17), 48–53.

(18) Nan, H.; Wang, Z.; Wang, W.; Liang, Z.; Lu, Y.; Chen, Q.; He, D.; Tan, P.; Miao, F.; Wang, X.; Wang, J.; Ni, Z. Strong Photoluminescence Enhancement of MoS2 through Defect Engineering and Oxygen Bonding. *ACS Nano* **2014**, *8* (6), 5738–5745.

(19) Qin, S.; Lei, W.; Liu, D.; Chen, Y. In-Situ and Tunable Nitrogen-Doping of MoS 2 Nanosheets. *Sci. Rep.* **2014**, *4*, 7582.

(20) Laskar, M. R.; Nath, D. N.; Ma, L.; Ii, E. W. L.; Lee, C. H.; Kent, T.; Yang, Z.; Mishra, R.; Roldan, M. A.; Idrobo, J.; Pantelides, S. T.; Pennycook, S. J.; Myers, C.; Wu, Y.; Rajan, S.; Idrobo, J.; Pantelides, S. T.; Pennycook, S. J. P-Type Doping of MoS2 Thin Films Using Nb P-Type Doping of MoS 2 Thin Films Using Nb. *Appl. Phys. Lett.* **2014**, *92104*, 104–108.

(21) Azcatl, A.; Qin, X.; Prakash, A.; Zhang, C.; Cheng, L.; Wang, Q.; Lu, N.; Kim, M. J.; Kim, J.; Cho, K.; Hinkle, C. L.; Appenzeller, J.; Wallace, R. M. Covalent Nitrogen Doping and Compressive Strain in MoS 2 by Remote N 2 Plasma Exposure. *Nano Lett.* **2016**, *16* (9), 5437–5443.

(22) Pallecchi, E.; Lafont, F.; Cavaliere, V.; Schopfer, F.; Mailly, D.; Poirier, W.; Ouerghi, A. High Electron Mobility in Epitaxial Graphene on 4H-SiC(0001) via Post-Growth Annealing under Hydrogen. *Sci. Rep.* **2014**, *4*, 4558.

(23) Robinson, J. A.; Hollander, M.; Labella, M.; Trumbull, K. A.; Cavalero, R.; Snyder, D. W. Epitaxial Graphene Transistors : Enhancing Performance via Hydrogen. *Nano Lett.* **2011**, *11*, 3875–3880.

(24) Henck, H.; Pierucci, D.; Ben Aziza, Z.; Silly, M. G.; Gil, B.; Sirotti, F.; Cassabois, G.; Ouerghi, A. Stacking Fault and Defects in Single Domain Multilayered Hexagonal Boron Nitride. *Appl. Phys. Lett.* **2017**, *110* (2), 23101.

(25) Yan, L.; Zheng, B.; Zhao, F.; Li, S.; Gao, X.; Xu, B.; Weiss, S.; Zhao, Y. Chem Soc Rev Chemistry and Physics of a Single Atomic Layer : Strategies and Challenges for Functionalization of Graphene and Graphene-Based Materials. *Chem. Soc. Rev.* **2012**, *41*, 97–114.

(26) Pallecchi, E.; Ridene, M.; Kazazis, D.; Mathieu, C.; Schopfer, F.; Poirier, W.; Mailly, D.; Ouerghi, A. Observation of the Quantum Hall Effect in Epitaxial Graphene on SiC(0001) with Oxygen Adsorption. *Appl. Phys. Lett.* **2012**, *100* (25), 253109.

(27) Riedl, C.; Coletti, C.; Iwasaki, T.; Zakharov, a. a.; Starke, U. Quasi-Free-Standing Epitaxial Graphene on SiC Obtained by Hydrogen Intercalation. *Phys. Rev. Lett.* **2009**, *103* (24), 246804.

(28) Henck, H.; Pierucci, D.; Chaste, J.; Naylor, C. H.; Avila, J.; Balan, A.; Silly, M. G.; Maria, C.; Sirotti, F.; Johnson, A. T. C.; Lhuillier, E.; Ouerghi, A. Electrolytic Phototransistor Based on Graphene-MoS2 van Der Waals P-N Heterojunction with Tunable Photoresponse P-N Heterojunction with Tunable Photoresponse. *Appl. Phys. Lett.* **2016**, *109*, 113103.

(29) Pierucci, D.; Henck, H.; Avila, J.; Balan, A.; Naylor, C. H.; Patriarche, G.; Dappe, Y. J.; Silly, M. G.; Sirotti, F.; Johnson, A. T. C.; Asensio, M. C.; Ouerghi, A. Band Alignment and Minigaps in Monolayer MoS2-Graphene van Der Waals Heterostructures. *Nano Lett.* **2016**, *16* (7), 4054–4061.

(30) Mattila, S.; Leiro, J. a.; Heinonen, M.; Laiho, T. Core Level Spectroscopy of MoS2. *Surf. Sci.* **2006**, *600* (24), 5168–5175.

(31) Kim, I. S.; Sangwan, V. K.; Jariwala, D.; Wood, J. D.; Park, S.; Chen, K.; Shi, F.; Ruiz-zepeda, F.; Ponce, A.; Jose-, M.; Dravid, V. P.; Marks, T. J.; Hersam, M. C.; Lincoln, J. Influence of Stoichiometry on the Optical and Electrical Properties of Chemical Vapor Deposition Derived MoS 2. *ACS Nano* **2014**, *8* (10), 10551–10558.

(32) Eda, G.; Yamaguchi, H.; Voiry, D.; Fujita, T.; Chen, M.; Chhowalla, M. Photoluminescence from Chemically Exfoliated MoS2. *Nano Lett.* **2011**, *11* (12), 5111–5116.

(33) Levasseur, a; Vinatier, P.; Gonbeau, D. X-Ray Photoelectron Spectroscopy: A Powerful Tool for a Better Characterization of Thin Film Materials. *Bull. Mater. Sci.* **1999**, *22* (3), 607–614.

(34) Baker, M. A.; Gilmore, R.; Lenardi, C.; Gissler, W. XPS Investigation of Preferential Sputtering of S from MoS 2 and Determination of MoS X Stoichiometry from Mo and S Peak Positions. *Appl. Surf. Sci.* **1999**, *150*, 255–262.

(35) Fleischauer, P. D.; Lince, J. R. A Comparison of Oxidation and Oxygen Substitution in MoS 2 Solid Film Lubricants. *Tribol. Int.* **1999**, *32*, 627–636.

(36) Liu, X.; Balla, I.; Bergeron, H.; Campbell, G. P.; Bedzyk, M. J.; Hersam, M. C. Rotationally Commensurate Growth of MoS2 on Epitaxial Graphene. *ACS Nano* **2015**, *10* (1), 1067–1075.

(37) Han, S. W.; Cha, G. B.; Frantzeskakis, E.; Razado-Colambo, I.; Avila, J.; Park, Y. S.; Kim, D.; Hwang, J.; Kang, J. S.; Ryu, S.; Yun, W. S.; Hong, S. C.; Asensio, M. C. Band-Gap Expansion in the Surface-Localized Electronic Structure of MoS 2(0002). *Phys. Rev. B* **2012**, *86*, 115105.

(38) Search, H.; Journals, C.; Contact, A.; Iopscience, M.; Phys, S. S.; Address, I. P. The Vibration Spectrum of Hydrogen Bound by Molybdenum Sulphide Catalysts. *J. Phys. C Solid State Phys.* **1981**, *14*, 4969–4983.

(39) Cristol, S.; Paul, J. F.; Payen, E.; Hutschka, F. Theoretical Study of the MoS 2 ( 100 ) Surface : A Chemical Potential Analysis of Sulfur and Hydrogen Coverage. *J. Phys. Chem. B* **2000**, *104*, 11220–11229.

(40) Spirko, J. A.; Neiman, M. L.; Oelker, A. M.; Klier, K. Electronic Structure and Reactivity of Defect MoS 2 II . Bonding and Activation of Hydrogen on Surface Defect Sites and Clusters. *Surf. Sci.* **2004**, *572*, 191–205.

(41) Coy Diaz, H.; Avila, J.; Chen, C.; Addou, R.; Asensio, M. C.; Batzill, M. Direct Observation of

Interlayer Hybridization and Dirac Relativistic Carriers in Graphene/MoS2 van Der Waals Heterostructures. *Nano Lett.* **2015**, *15* (2), 1135–1140.

(42)  Pierucci, D.; Henck, H.; Naylor, C. H.; Sediri, H.; Lhuillier, E.; Balan, A.; Rault, J. E.; Dappe, Y. J.; Bertran, F.; Le Févre, P.; Johnson, A. T. C.; Ouerghi, A. Large Area Molybdenum Disulphide-Epitaxial Graphene Vertical Van Der Waals Heterostructures. *Sci. Rep.* **2016**, *6*, 26656.

(43)  Gierz, I.; Henk, J.; Hochst, H.; Ast, C. R.; Kern, K. Illuminating the Dark Corridor in Graphene: Polarization Dependence of Angle-Resolved Photoemission Spectroscopy on Graphene. *Phys. Rev. B - Condens. Matter Mater. Phys.* **2011**, *83* (12), 1–4.

(44)  Wang, Z.; Chen, Q.; Wang, J. Electronic Structure of Twisted Bilayers of graphene/MoS2 and MoS2/MoS2. *J. Phys. Chem. C* **2015**, *119* (9), 4752–4758.

(45)  Ben Aziza, Z.; Henck, H.; Di, D.; Pierucci, D.; Chaste, J.; Naylor, C. H.; Balan, A.; Dappe, Y. J.; Johnson, A. T. C.; Ouerghi, A. Bandgap Inhomogeneity of MoS 2 Monolayer on Epitaxial Graphene Bilayer in van Der Waals P-N Junction. *Carbon N. Y.* **2016**, *110*, 396–403.

(46)  Li, H.; Zhang, Q.; Yap, C. C. R.; Tay, B. K.; Edwin, T. H. T.; Olivier, A.; Baillargeat, D. From Bulk to Monolayer MoS 2: Evolution of Raman Scattering. *Adv. Funct. Mater.* **2012**, *22* (7), 1385–1390.

(47)  Lee, C.; Yan, H.; Brus, L. E.; Heinz, T. F.; Hone, Ḱ. J.; Ryu, S. Anomalous Lattice Vibrations of Single- and Few-Layer MoS2. *ACS Nano* **2010**, *4* (5), 2695–2700.

(48)  Han, S. W.; Yun, W. S.; Lee, J. D.; Hwang, Y. H.; Baik, J.; Shin, H. J.; Lee, W. G.; Park, Y. S.; Kim, K. S. Hydrogenation-Induced Atomic Stripes on the 2 H -MoS 2 Surface. *Phys. Rev. B.* **2015**, *241303*, 1–5.

(49)  Wang, Y.; Cong, C.; Qiu, C.; Yu, T. Raman Spectroscopy Study of Lattice Vibration and Crystallographic Orientation of Monolayer mos2 under Uniaxial Strain. *Small* **2013**, *9* (17), 2857–2861.

(50)  Hui, Y. Y., Liu, X., Jie, W., Chan, N. Y., Hao, J., Hsu, Y. T., Li, L. J., Guo W., Lau, S. P. Exceptional Tunability of Band Energy in a Compressively Strained Trilayer. *ACS Nano* **2013**, *7* (8), 7126–7131.

(51)  Nayak, A. P.; Bhattacharyya, S.; Zhu, J.; Liu, J.; Wu, X.; Pandey, T.; Jin, C.; Singh, A. K.; Akinwande, D.; Lin, J. Transition in Multilayered Molybdenum Disulphide. *Nat. Commun.* **2014**, *5*, 3731.

(52)  Chakraborty, B.; Bera, A.; Muthu, D. V. S.; Bhowmick, S.; Waghmare, U. V.; Sood, A. K. Symmetry-Dependent Phonon Renormalization in Monolayer MoS 2 Transistor. *Phys. Rev. B - Condens. Matter Mater. Phys.* **2012**, *85* (16), 2–5.

(53)  Kim, B. H.; Park, M.; Lee, M.; Baek, S. J.; Jeong, H. Y.; Choi, M.; Chang, S. J.; Hong, W. G.; Kim, T. K.; Moon, H. R.; Park, W.; Park, N.; Jun, Y. RSC Advances Structure of MoS2 Induced by Molecular Hydrogen Treatment at Room Temperature. *RSC Adv.* **2013**, *3*, 18424–18429.

(54)  Han, G. H.; Kybert, N. J.; Naylor, C. H.; Lee, B. S.; Ping, J.; Park, J. H.; Kang, J.; Lee, S. Y.; Lee, Y. H.; Agarwal, R.; Johnson,  a T. C. Seeded Growth of Highly Crystalline Molybdenum Disulphide

Monolayers at Controlled Locations. *Nat. Commun.* **2015**, *6*, 6128.

(55) Polack, F.; Silly, M.; Chauvet, C.; Lagarde, B.; Bergeard, N.; Izquierdo, M.; Chubar, O.; Krizmancic, D.; Ribbens, M.; Duval, J.-P.; Basset, C.; Kubsky, S.; Sirotti, F.; Garrett, R.; Gentle, I.; Nugent, K.; Wilkins, S. TEMPO: A New Insertion Device Beamline at SOLEIL for Time Resolved Photoelectron Spectroscopy Experiments on Solids and Interfaces. *AIP Conf. Proc* **2010**, *1234* (2010), 185–188.

(56) Lewis, J. P.; Jelínek, P.; Ortega, J.; Demkov, A. A.; Trabada, D. G.; Haycock, B.; Wang, H.; Adams, G.; Tomfohr, J. K.; Abad, E.; Wang, H.; Drabold, D. A. Advances and Applications in the FIREBALL Ab Initio Tight-Binding Molecular-Dynamics Formalism. *Phys. Status Solidi Basic Res.* **2011**, *248* (9), 1989–2007.

(57) Jelínek, P.; Wang, H.; Lewis, J.; Sankey, O.; Ortega, J. Multicenter Approach to the Exchange-Correlation Interactions in Ab Initio Tight-Binding Methods. *Phys. Rev. B* **2005**, *71*, 235101.

# Tunable Doping in Hydrogenated Single Layered Molybdenum Disulfide

Debora Pierucci[1], Hugo Henck[1], Zeineb Ben Aziza[1], Carl H. Naylor[2], A. Balan[2], Julien E. Rault[3], M. G. Silly[3], Yannick J. Dappe[4], François Bertran[3], Patrick Le Fevre[3], F. Sirotti[3], A.T Charlie Johnson[2] and Abdelkarim Ouerghi[1*]

[1]Centre de Nanosciences et de Nanotechnologies, CNRS, Univ. Paris-Sud, Université Paris-Saclay, C2N – Marcoussis, 91460 Marcoussis, France
[2]Department of Physics and Astronomy, University of Pennsylvania, 209S 33rd Street, Philadelphia, Pennsylvania 19104, USA
[3] Synchrotron-SOLEIL, Saint-Aubin, BP48, F91192 Gif sur Yvette Cedex, France
[4] SPEC, CEA, CNRS, Université Paris-Saclay, CEA Saclay, 91191 Gif-sur-Yvette Cedex, France

*Corresponding author, E-mail: abdelkarim.ouerghi@lpn.cnrs.fr

1. **XPS**

The C 1s spectrum for a p-doped quasi-freestanding pristine bilayer graphene is shown in Figure S1 (bottom). Only two components are present on the spectra[1] due to the bilayer graphene ( G peak at binding energy BE = 284.3 eV) and the SiC substrate (BE = 282.5 eV). Respect to an as grown n-doped monolayer epitaxial graphene[2] the G peak presents a shift of about 0.4 eV to lower BE indicating a change in the doping, from n to p, induced by the hydrogenation process[1]. Moreover, the SiC component is shifted of about 1 eV to lower BE, which confirms that the hydrogen bonds are present at the SiC surface inducing this band bending variation confirming a complete decoupling of the buffer layer. The spectrum presents no change after the three hydrogen doses (~ $5 \times 10^2$ L, $5 \times 10^3$ L and $8 \times 10^4$ L, where 1 Langmuir (L) = $10^{-6}$ Torr $\times$ s).

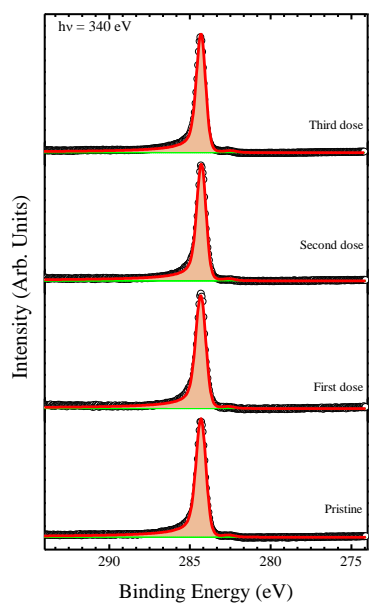

**Figure S1:** High-resolution C 1s spectrum (hv = 340 eV) of pristine graphene and graphene under three differ hydrogen doses ((~ $5 \times 10^2$ L, $5 \times 10^3$ L and $8 \times 10^4$ L). The experimental data points are displayed as dots. The solid line is the envelope of fitted components.

## 2. ARPES

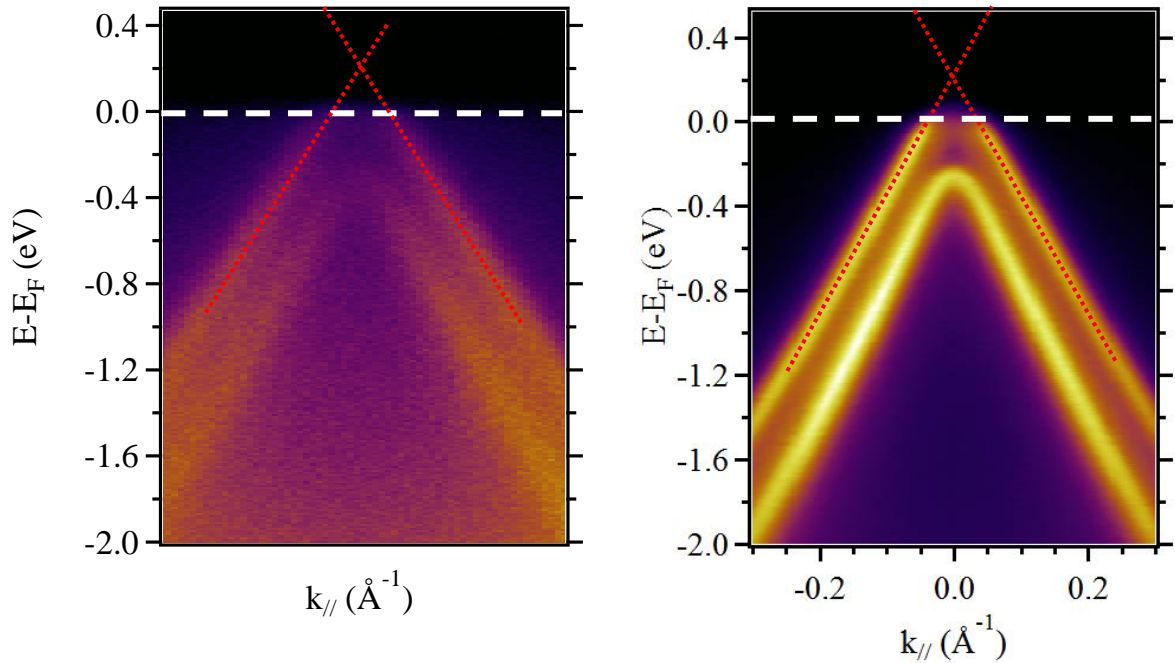

**Figure S2:** a) and b) ARPES at room temperature of MoS$_2$/graphene and hydrogenated MoS$_2$/graphene heterostructure, measured at hν = 100 eV and hν = 50 eV respectively, around the K-point and along the graphene ΓK direction.

## 3. PL:

Figure S3 shows the photoluminescence spectra of MoS$_2$ direct after its transfer on p-doped graphene and after a hydrogenation process. The decrease of the intensity before and after hydrogenation can be explained by the suppression of defects in the MoS$_2$ as evidenced elsewhere.[3]

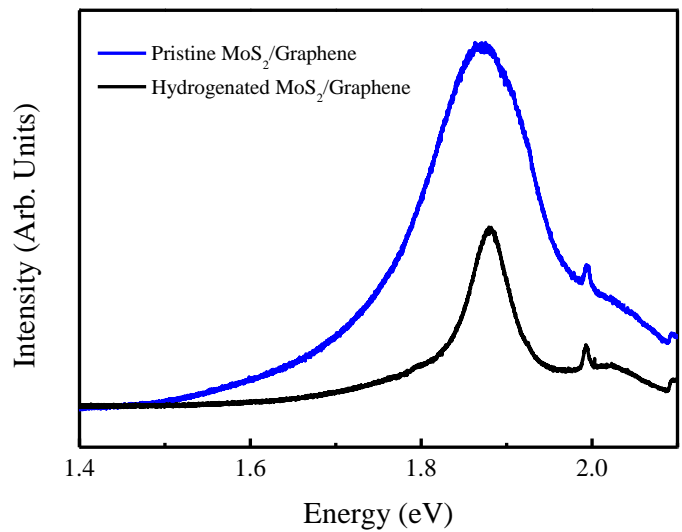

**Figure S3:** Photoluminescence spectra of pristine and hydrogenated MoS$_2$

**References:**

(1) Riedl, C.; Coletti, C.; Iwasaki, T.; Zakharov, a. a.; Starke, U. Quasi-Free-Standing Epitaxial Graphene on SiC Obtained by Hydrogen Intercalation. *Phys. Rev. Lett.* **2009**, *103* (24), 246804.

(2) Velez-fort, E.; Mathieu, C.; Pallecchi, E.; Pigneur, M.; Silly, M. G.; Belkhou, R.; Marangolo, M.; Shukla, A.; Sirotti, F.; Ouerghi, A.; Curie, M.; Cedex, Y. Epitaxial Graphene on 4H-SiC ( 0001 ) Grown under Nitrogen Flux : Evidence of Low Nitrogen Doping and High Charge Transfer. *ACS Nano* **2012**, *6* (12), 10893–10900.

(3) Parkin, W. M.; Balan, A.; Liang, L.; Das, P. M.; Lamparski, M.; Naylor, C. H.; Rodr, J. A.; Johnson, A. T. C.; Meunier, V.; Drndic, M. Raman Shifts in Electron-Irradiated Monolayer. **2016**.